\documentstyle[twoside,fleqn,espcrc2,epsf]{article}

\def\spose#1{\hbox to 0pt{#1\hss}}
\def\ltapprox{\mathrel{\spose{\lower 3pt\hbox{$\mathchar"218$}}
 \raise 2.0pt\hbox{$\mathchar"13C$}}}
\def\gtapprox{\mathrel{\spose{\lower 3pt\hbox{$\mathchar"218$}}
 \raise 2.0pt\hbox{$\mathchar"13E$}}}
\def\inapprox{\mathrel{\spose{\lower 3pt\hbox{$\mathchar"218$}}
 \raise 2.0pt\hbox{$\mathchar"232$}}}

\newcommand{\beq}{\begin{equation}}
\newcommand{\eeq}{\end{equation}}
\newcommand{\beqn}{\begin{eqnarray}}
\newcommand{\eeqn}{\end{eqnarray}}

\title{Parallel computing and molecular dynamics 
of biological membranes}

\author{G.~La Penna, S.~Letardi, V.~Minicozzi, S.~Morante,
G.C.~Rossi and G.~Salina\\
Presented by G.C.~Rossi\address{Dip. di Fisica, Universit\`a di Roma 
``{\it Tor Vergata}''
and I.N.F.N., Sezione di Roma II, \\ 
Via della Ricerca Scientifica 1, I-00133 Roma - Italy}}

\begin{document}

\begin{abstract}

In this talk I discuss the general question of the portability of Molecular 
Dynamics codes for diffusive systems on parallel computers of the 
APE family. The intrinsic single precision arithmetics
of the today available APE platforms does not seem to affect the numerical 
accuracy of the simulations, while the absence of integer addressing from
CPU to individual nodes puts strong constraints on the possible programming 
strategies. Liquids can be very satisfactorily simulated using
the ``systolic" method. For more complex 
systems, like the biological ones at which we are ultimately interested in, 
the ``domain decomposition" approach is best suited to beat the quadratic
growth of the inter-molecular computational time with
the number of elementary components of the system. The promising perspectives
of using this strategy for extensive simulations of lipid bilayers are 
briefly reviewed.
\end{abstract}
\maketitle

\section{Introduction}
In simulating the behaviour of microscopic systems
Molecular Dynamics (MD) faces two major problems. One is the intrinsic 
limitation coming from the  use of classical mechanics to describe the 
dynamics of the ``elementary" 
components of the system~\cite{AT}. The second, not less important, is of 
practical nature and it has to do with the finiteness of computer resources
available at any time.
In recent investigations~\cite{LPMMRS}~\cite{NEW} we addressed 
the second of these questions, showing that parallel computers,
particularly of the APE family~\cite{APE}, can be successfully employed to 
speed up in a substantial way MD simulations of diffusive systems, i.e.~of 
systems in which the ``list" of atoms that have a non-negligible interaction 
with a given atom of the system changes with time. In a solid (or in 
Lattice Quantum Chromo-Dynamics - LQCD) the 
``list" is blocked and is assigned once for all at the beginning of the 
simulation.

Static and dynamic properties of liquids can be adequately 
studied~\cite{LPMMRS}, using the ``systolic" method~\cite{PET}. For 
the more complex case of lipid bilayers we have developed new approaches,
adapting to these systems the general ``domain decomposition" strategy. The
latter has the virtue of leading to CPU times for the computation of the 
inter-molecular potential that (at constant density) 
grow only linearly with the number of atoms.
\section{Liquid Butane}
As a first significant test case, we have studied in great detail liquid 
butane ($C_4 H_{10}$)~\cite{LPMMRS}, using a standard Multiple Time Step (MTS)
integration algorithm~\cite{AT} (with a ``long" integration time step 
$\Delta t_L= 4$~$fs$ and a partition number $P_{\circ}=8$). The results of the 
simulation of the time evolution of a system of $M=512$ molecules confirm that 
for a homogeneous diffusive system the ``systolic" method~\cite{PET} is well 
suited for massive parallel production. The method consists in computing the 
inter-molecular (Lennard-Jones) interactions between atoms belonging to 
different molecules, by first democraticaly distributing in bunches among 
the $N$ nodes of the machine the molecules of the system.
Coordinates and momenta of the $N$ bunches of molecules are copied in 
transitory arrays and circulated through nodes.
By bringing them successively in contact with the node-residing molecules, 
the crossed interaction terms are all computed in $N-1$ moves.
The method does not beat the $M^2$ growth of the inter-molecular computational 
time, but decreases it by a factor equal to the number of nodes.

On the 512-nodes APE configuration ({\it {Torre}}) it took in all 450 hours of 
CPU time to collect the whole statistics of 10~$ns$ presented in~\cite{LPMMRS}. 
Computer times of this size are certainly within the current standards of MD 
and LQCD simulations. 
For comparison, a system of 256 molecules has been simulated, in double 
precision, on a DIGITAL 200 4/233 $\alpha$-station. To give an approximate 
reference figure we may quote a factor of 50, as a gain in speed in going from 
the $\alpha$-station to the {\it {Torre}}. 

Our code was written in TAO, the APE highest level language. However APE 
simulation times can be substantially reduced (by a factor from 3 to 4), 
if the APE-{\it assembler} micro-code of the most time-consuming part of the 
program, where the inter-molecular forces are computed, is properly optimized.
\begin{figure}[htb]
\vspace{2.5cm}
\includegraphics{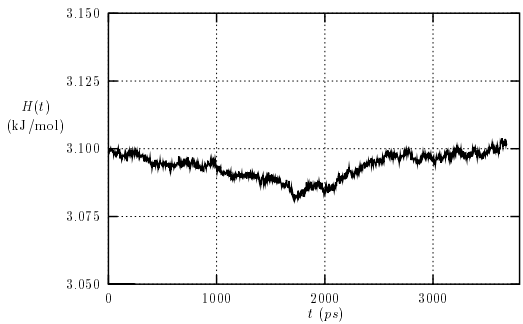}
\null\vskip 0.4cm
\caption{The time evolution of the total energy of a system of 512 molecules 
of liquid butane.}
\label{fig:H}
\end{figure}

Very good agreement is obtained among the various simulations. Our results also 
compare nicely with the available experimental 
data~\cite{HOU}~\cite{PCEH}~\cite{HCP} and with the 
pioneering simulation of ref.~\cite{RB}, where 64 molecules of butane were 
followed for up to 20~$ps$ (after few $ps$ of equilibration).


In our very long simulations we did 
not see signs of instabilities nor of any systematic drift in the value of the 
total energy of the system (see Fig.~\ref{fig:H}). 
Single precision rounding problems were sometimes held responsible for 
pathologies of that kind.
\section{Towards simulating realistic membranes} 
Simulating realistic cell membranes and studying their 
interaction with small peptides (mimicking pharmaceuticals of possible 
therapeutical interest) is of the utmost importance, if not for 
immediate medical use, certainly for the development of new conceptual and 
practical tools in MD applications to biological systems. A lot of work 
has gone in this direction (see e.g.~\cite{BEB}~\cite{HSS} and 
references therein), but we are still far from having a viable tool-kit
for immediate practical use.

Schematically a cell membrane is constituted by an almost spherical bilayer of
phospho-lipidic molecules, separating the interior of the cell from the 
external world. Various kinds of peptides and 
proteins, responsible for the biochemical processes necessary 
for the life and the functionality of the cell,
are plugged into the membrane.

Phospho-lipids are large Y-shaped molecules (with more than 30 atoms, 
not counting carbon bound hydrogens) with a hydrophilic head and two 
hydrophobic tails. This hydropathicity configuration lead to a well
defined bilayer 3-D structure: the hydrophilic 
heads are in contact with water, present both outside
and inside the cell, while the hydrophobic tails are more or less 
back-to-back pair-wise aligned.

An important parameter governing the reaction rate of many 
biological processes taking place in the membrane is its ``permeability".
In this respect a membrane can be regarded as a liquid crystal,
with a permeability which depends ``critically" on the temperature, on the 
detailed chemical composition of the constituent phospho-lipids and on the
concentration of chemicals dispersed in the membrane itself or in the solvent. 

Even from this very crude picture, it is clear that a detailed simulation 
of the dynamics of the membrane of a living cell is just 
impossible and we have to resort to a number of simplifications. As it appears 
experimentally that the nature and the location of the phase transitions, which
control the physico-chemical properties of the membrane, are related to the 
bulk ordering properties of the hydrophobic tails, a first step in the 
direction of simulating a realistic system is to take a sufficiently large 
bilayer in a acqueous medium and begin to study 1) the behaviour of the 
relevant order parameters as functions of the temperature, 2) the dependence 
of the position of critical points upon the concentration of small 
intramembrane peptides.

We have started our investigation  
with a system of $2\times 256$ Dimyristoyl-phosphatidylcholine (DMPC) 
molecules (each molecule is composed by 37 atoms) in vacuum, neglecting 
in these first trial simulations Coulomb 
interactions\footnote{Special care must be exerced 
to deal with Coulomb interaction. I will not discuss this problem here. 
The interested reader can have a look to~\cite{TB} and references 
therein.}. We have run the dynamics of the system at various 
temperatures on the {\it Torre}. In each run the history of the system
was followed for several hundreds $ps$ (plus equilibration). At very low 
temperatures ($T<200$ K) the system appears to be stable, 
although we know that, lacking Coulomb interactions and in absence of solvent, 
it is actually unstable and expected to ``explode" at higher $T$. 

Already in this oversimplified test case CPU simulation times are exceedingly 
large, as the computation of the inter-molecular forces require the 
evaluation of a daring $(2\times 256 \times 37)^2/2$ terms!
Since the vast majority of them gives a negligibly 
small contribution to the forces (the Lennard-Jones potential decreases very 
fast with the distance), the obvious way to cope with this problem is to
avoid computing the very many effectively irrelevant terms. To this end the 
physical space in which the system lives is first decomposed 
into $N$ domains, each one attributed to one of the nodes of the machine. 
The domains are in turn subdivided in cells. The number of cells in each domain 
and, hence, the spatial extension of each cell, is chosen so that 
the interaction between atoms residing in non-nearest-neighboring cells is 
negligibly small. Then inter-molecular interactions are computed only between
the atoms of a cell and those of the 26/2 nearest neighboring ones. Every 
$N_{remap}$ integration steps, the coordinates of all the molecules 
are cross-checked node by node and, if necessary, molecules that have
wandered away from the original cell are reassigned to the cell to which  
they came up to belong. The remapping of the system costs a 
time which only growth linerly with the number of particles.

A problem with this approach on APE platforms is the lack 
of integer addressing to individual nodes, which makes impossible to assign 
locally to each node the set of indices representing the number of 
interaction terms to be computed between pairs of cells. 
This difficulty has been overcome by assigning to all nodes the same set of 
indices, namely the set of the largest values taken by them throughout the
machine. Nodes with fewer than the maximal number of terms to be computed will 
wait until other nodes have finished their job.
\begin{figure}[htb]
\vspace{2.8cm}	
\includegraphics{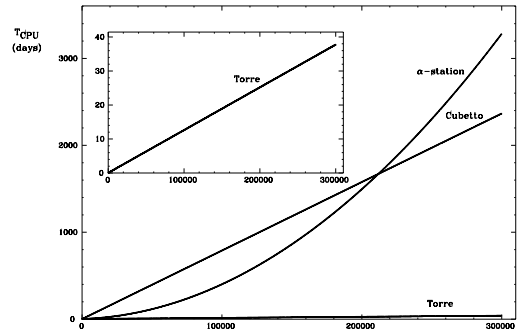}
\null\vskip 0.4cm
\caption{$\alpha$-station, {\it Cubetto} and {\it Torre} 
CPU times}
\label{fig:times}
\end{figure}

We have fully implemented these ideas on APE computers in the case of butane, 
obtaining the expected linear behaviour with the number, $n$,  of interacting 
particles and a perfect scalability with the number of nodes
in going from the {\it Cubetto} ($N=2^3$) to the {\it Torre} ($N=2^9$). 
CPU times for simulating the dynamics of a system of 2048 
molecules of butane for 1~$ns$ with our standard MTS algorithm were measured 
on the Digital 200 4/233 $\alpha$-station, the {\it Cubetto} and the 
{\it Torre}, obtaining
\begin{equation}
\begin{array}{l}
T^{\alpha}_{_{CPU}}=2.3\cdot  10^{-4}[2.3 n + \frac{6 \cdot 10^{-3}}
{N_{update}} n^2] \,\,\,\;\quad {\mbox{days}}\\
T^{Cubetto}_{_{CPU}}=2.3\cdot 10^{-3}[3.4 n + \frac{0.22}{N_{remap}} n]
\quad {\mbox{days}}\\
T^{Torre}_{_{CPU}}=2.3 \cdot 10^{-5}[5.4 n + \frac{1.4}{N_{remap}} n]
\,\,\,\;\quad {\mbox{days}}
\end{array}
\label{T}
\end{equation}

Eqs.~\ref{T} with $N_{remap}=N_{update}=40$ are plotted in 
Fig.~\ref{fig:times} as functions of $n$. 
Notice that in the case of the $\alpha$-station we have used the 
``list" method to speed up the simulation. As the list 
updating time growths quadratically with $n$, 
for a sufficiently large number of atoms the ``list"
method curve will always cross the ``domain decomposition" straight line. 
\vskip .2cm

{\bf Acknowledgments}

Partial support from CNR and INFM (Italy) and from the EC contract
CHRX-CT92-0051 is acknowledged.

\end{document}